DFTT 12/94

# A Model Independent Approach to Future Solar Neutrino Experiments

S.M. Bilenky[a,b,c]* and C. Giunti[b,c]⋆

[a] Joint Institute of Nuclear Research, Dubna, Russia

[b] INFN Torino, Via P. Giuria 1, I–10125 Torino, Italy

[c] Dipartimento di Fisica Teorica, Università di Torino

## Abstract

We discuss here what model independent information about properties of neutrinos and of the sun can be obtained from future solar neutrino experiments (SNO, Super-Kamiokande). It is shown that in the general case of transitions of solar $\nu_e$'s into $\nu_\mu$ and/or $\nu_\tau$ the initial $^8$B neutrino flux can be measured by the observation of NC events. From the CC measurements the $\nu_e$ survival probability can be determined as a function of neutrino energy. The general case of transitions of solar $\nu_e$'s into active as well as sterile neutrinos is considered. A number of relations between measurable quantities the test of which will allow to answer the question whether there are sterile neutrinos in the solar neutrino flux on the earth are derived. Transitions of solar $\nu_e$'s into active and sterile states due to neutrino mixing and Dirac magnetic moments or into active left-handed neutrinos and active right-handed antineutrinos due to neutrino mixing and Majorana transition magnetic moments are also considered. It is shown that future solar neutrino experiments will allow to distinguish between the cases of Dirac and Majorana magnetic moments.

* BILENKY@TO.INFN.IT
⋆ GIUNTI@TO.INFN.IT

# 1 Introduction

The experiments on the detection of solar neutrinos are of great interest from the point of view of the investigation of neutrino properties as well as the investigation of the sun. These experiments are very sensitive to the parameters that characterize the Pontecorvo neutrino mixing: they allow to investigate a region of very small values of $\Delta m^2$ ($\Delta m^2 \equiv m_2^2 - m_1^2$ and $m_1$, $m_2$ are neutrino masses) and (if there are resonant MSW transitions in matter [1]) a wide region of mixing angles, including the theoretically important region of small angles. The observation of a periodical time variation of the solar neutrino flux in anticorrelation with the solar activity would be an evidence in favor of large neutrino magnetic moments ($\mu \sim 10^{-11} - 10^{-10}\mu_B$) [2, 3].

The data of present day experiments give some indications in favor of neutrino mixing (see ref.[4]) [1]. It is necessary however to stress that from the data of present day experiments it is possible to extract informations about neutrino mixing only by assuming that the initial neutrino fluxes are given by the Standard Solar Model (SSM) [6, 7]. There are many input parameters that are used in the SSM, such as the cross sections of the different nuclear reactions, opacities and others. The values of the cross sections were obtained by a delicate procedure of extrapolation of the existing experimental data to low energies. The usual method of extrapolation has been recently criticized and a smaller value for the $^8$B solar neutrino flux was obtained in the framework of a new SSM [8].

In refs.[9, 10] we discussed which kind of model independent information about Pontecorvo neutrino mixing and the solar neutrino flux can be extracted from the real-time SNO [11] and Super-Kamiokande [12] experiments (scheduled to start in 1995 and 1996, respectively). In this paper we will continue the development of this model independent approach to future solar neutrino experiments.

The new feature of the SNO experiment will be the measurement of the energy spectrum of solar $\nu_e$ on the earth by the observation of the CC process

$$\nu_e + d \to e^- + p + p \tag{1.1}$$

and the simultaneous measurement of the total rate of NC events

$$\nu_x + d \to \nu_x + p + n \tag{1.2}$$

with $x = e, \mu, \tau$. Solar neutrinos will be detected in this experiment also through the observation ($\simeq$ 1 event/day) of the ES process

$$\nu_x + e^- \to \nu_x + e^- \tag{1.3}$$

In the Super-Kamiokande experiment (S-K) solar neutrinos will be detected through the observation of ES events which much higher statistics ($\simeq$ 30 event/day).

We will show that, if solar $\nu_e$ can transform only into $\nu_\mu$ and/or $\nu_\tau$, these future solar neutrino experiments will allow:

[1] However, they can be also explained by neutrino spin-flavor precession in solar matter due to neutrino masses and magnetic moments (see ref.[5]).

(1) To *measure* the total flux of initial $^8$B neutrinos.

(2) To determine directly from the experimental data the $\nu_e$ survival probability as a function of neutrino energy, which give a model independent information about the mechanism of transition of solar $\nu_e$'s into other states [2].

In the general case of neutrino mixing solar $\nu_e$'s can transform not only into active neutrinos, but also into left-handed sterile states. We will show that future solar neutrino experiments will allow to check in a model independent way whether there are such transitions. We will also obtain lower bounds for the probability of transition of solar $\nu_e$'s into all possible sterile states. Let us stress the importance of the problem of existence of sterile neutrinos. Dirac neutrino masses could be generated in the framework of the Standard Model with the help of the standard Higgs mechanism (through a Higgs doublet). Therefore transitions between active neutrino states could be an effect of the Standard Model. A detection of transitions into sterile states would be a discovery of an effect beyond the Standard Model.

In the last part of this paper we will discuss possibilities of a model independent treatment of the data from future solar neutrino experiments in the case of spin and/or spin-flavor precession of solar $\nu_e$'s due to anomalously large neutrino magnetic moments. The effects of neutrino magnetic moments [2, 3] were widely discussed in last years (see ref.[13]) in connection with a possible indication of the existence of an anticorrelation between the flux of solar $\nu_e$'s and the sunspot number [14].

The transitions of neutrinos in the magnetic field of the sun depend on the nature of neutrinos. In the case of Dirac neutrinos, solar $\nu_e$'s can be transformed into sterile right-handed neutrinos $\nu_{\ell R}$ ($\ell = e, \mu, \tau$), quanta of right-handed fields. In the case of Majorana neutrinos, transitions of solar $\nu_e$'s into active right-handed antineutrinos $\bar{\nu}_\ell$ ($\ell = e, \mu, \tau$), quanta of left-handed fields, could take place. We will show that in the case of Majorana transition magnetic moments the NC event rate $N^{\rm NC}$ does not depend on time and a measurement of $N^{\rm NC}$ will allow to determine the initial flux of $^8$B neutrinos with a theoretical uncertainty of a few %. In the case of Dirac magnetic moments the NC event rate will depend on time and we obtain model independent lower bounds for the transition probability of initial $\nu_e$'s into right-handed sterile neutrinos.

## 2 Transitions of solar $\nu_e$'s into $\nu_\mu$ and/or $\nu_\tau$

In this section we will consider the possibility of a model independent treatment of the data from the SNO and S-K experiments in the general case of transitions of solar $\nu_e$'s into $\nu_\mu$ and $\nu_\tau$. In these experiments, due to the high energy threshold ($\simeq 6\,$MeV for the CC process, $\simeq 5\,$MeV for the ES process and $2.2\,$MeV for the NC process), only neutrinos coming from

[2]The set of measurements that determines the initial $\nu_e$ flux and the probability of $\nu_e$'s to survive as a function of neutrino energy can naturally be called a complete set of solar neutrino measurements.

$^8$B decay will be detected. The distortions of the neutrino spectra are negligibly small under solar conditions [15] and the energy spectrum of the initial $^8$B $\nu_e$'s is given by

$$\phi_{\nu_e}^{^8\mathrm{B}}(E) = \Phi_{\nu_e}^{^8\mathrm{B}}\, X(E) \tag{2.1}$$

The function $X(E)$ is the normalized $\left(\int X(E)\,\mathrm{d}E = 1\right)$ neutrino spectrum from the decay $^8\mathrm{B} \to {}^8\mathrm{Be} + e^+ + \nu_e$, which is determined by the phase space factor (small corrections due to forbidden transitions where calculated in Ref.[16]). The factor $\Phi_{\nu_e}^{^8\mathrm{B}}$ in Eq.(2.1) is the total flux of initial $^8$B solar $\nu_e$'s.

Using only the $\nu_e$, $\nu_\mu$, $\nu_\tau$ universality of NC, for the total flux of initial $^8$B $\nu_e$'s we obtain

$$\Phi_{\nu_e}^{^8\mathrm{B}} = \frac{N^{\mathrm{NC}}}{X_{\nu d}^{\mathrm{NC}}} \tag{2.2}$$

Here $N^{\mathrm{NC}}$ is the total NC event rate in the SNO experiment and

$$X_{\nu d}^{\mathrm{NC}} \equiv \int_{E_{\mathrm{th}}^{\mathrm{NC}}} \sigma_{\nu d}^{\mathrm{NC}}(E)\, X(E)\, \mathrm{d}E \tag{2.3}$$

where $\sigma_{\nu d}^{\mathrm{NC}}(E)$ is the cross section for the process $\nu\, d \to \nu\, n\, p$. Using the results of the calculation of $\sigma_{\nu d}^{\mathrm{NC}}(E)$ presented in ref.[17], we obtained [3] $X_{\nu d}^{\mathrm{NC}} = 4.72 \times 10^{-43}\,\mathrm{cm}^2$. Therefore, the measurement of the NC event rate in the SNO experiment will allow to determine the initial flux of $^8$B neutrinos independently from possible transitions of solar $\nu_e$'s into $\nu_\mu$ and/or $\nu_\tau$, i.e. *to study the sun independently from the properties of neutrinos.*

In the SNO experiment the electron spectrum $n^{\mathrm{CC}}(T_e)$ of the CC reaction (1.1) will be measured and the spectrum of solar $\nu_e$ on the earth $\phi_{\nu_e}(E)$ will be determined ($T_e$ is the electron kinetic energy and $E$ is the neutrino energy). In the following we will consider $\phi_{\nu_e}(E)$ as a measurable quantity. From Eqs.(2.1) and (2.2) for the $\nu_e$ survival probability we have

$$\mathrm{P}_{\nu_e \to \nu_e}(E) = \alpha(E)\, \frac{\phi_{\nu_e}(E)}{N^{\mathrm{NC}}} \tag{2.4}$$

where

$$\alpha(E) \equiv \frac{X_{\nu d}^{\mathrm{NC}}}{X(E)} \tag{2.5}$$

is a known quantity. Thus, the $\nu_e$ survival probability will be determined directly from the NC and CC data of the SNO experiment and consequently *the properties of neutrinos will be studied by this experiment independently from the SSM.*

Using Eqs.(2.2) for the total rate of ES events we have:

$$N^{\mathrm{ES}} = \int_{E_{\mathrm{th}}^{\mathrm{ES}}} \left[\sigma_{\nu_e e}(E) - \sigma_{\nu_\mu e}(E)\right] \phi_{\nu_e}(E)\, \mathrm{d}E + \frac{X_{\nu_\mu e}^{\mathrm{ES}}}{X_{\nu d}^{\mathrm{NC}}}\, N^{\mathrm{NC}} \tag{2.6}$$

[3] We used the values of the function $X(E)$ given in ref.[6].

where $\sigma_{\nu_\ell e}(E)$ is the cross section of the process $\nu_\ell \, e^- \to \nu_\ell \, e^-$ $(\ell = e, \mu)$ and

$$X^{\rm ES}_{\nu_\mu e} \equiv \int_{E^{\rm ES}_{\rm th}} \sigma_{\nu_\mu e}(E) \, X(E) \, {\rm d}E \tag{2.7}$$

For $T_e^{\rm th} = 4.5\,{\rm MeV}$ we obtained $X^{\rm ES}_{\nu_\mu e} = 3.23 \times 10^{-45}\,{\rm cm}^2$ (in this calculation we used $g_V = -0.036$ and $g_A = -0.505$ [18]).

Thus the measurement of $N^{\rm NC}$ and $n^{\rm CC}(T_e)$ in the SNO experiment will allow to *predict* the total ES event rate in the SNO and S-K experiments. It is possible to show that these measurements will allow to predict not only the total ES event rate, but also the *spectrum* of the recoil electrons in the process $\nu e^- \to \nu e^-$. In fact, using Eqs.(2.2) for the spectrum of the recoil electrons we have

$$n^{\rm ES}(T_e) = \int_{E_{\rm m}(T_e)} \left[ \frac{{\rm d}\sigma_{\nu_e e}}{{\rm d}T_e}(E,T_e) - \frac{{\rm d}\sigma_{\nu_\mu e}}{{\rm d}T_e}(E,T_e) \right] \phi_{\nu_e}(E) \, {\rm d}E + \frac{X^{\rm ES}_{\nu_\mu e}(T_e)}{X^{\rm NC}_{\nu d}} \, N^{\rm NC} \tag{2.8}$$

Here $T_e$ is the kinetic energy of the recoil electrons,

$$\frac{{\rm d}\sigma_{\nu_\ell e}}{{\rm d}T_e}(E,T_e) = \frac{2\,{G_{\rm F}}^2\,m_e}{\pi} \left[ (g_L^{\nu_\ell})^2 + (g_R^{\nu_\ell})^2 \left(1 - \frac{T_e}{E}\right)^2 - g_L^{\nu_\ell}\,g_R^{\nu_\ell}\,\frac{m_e\,T_e}{E^2} \right] \tag{2.9}$$

is the differential cross section of the process $\nu_\ell e^- \to \nu_\ell e^-$ $(\ell = e, \mu)$, and

$$E_{\rm m}(T_e) = T_e \left( \frac{1}{2} + \frac{1}{2} \sqrt{1 + \frac{2\,m_e}{T_e}} \right) \tag{2.10}$$

In the Standard Model $g_L^{\nu_e} = 1/2 + \sin^2\theta_{\rm W}$, $g_L^{\nu_\mu} = -1/2 + \sin^2\theta_{\rm W}$ and $g_R^{\nu_e} = g_R^{\nu_\mu} = \sin^2\theta_{\rm W}$. Finally, $X^{\rm ES}_{\nu_\mu e}(T_e)$ is defined as

$$X^{\rm ES}_{\nu_\mu e}(T_e) \equiv \int_{E_{\rm m}(T_e)} \frac{{\rm d}\sigma_{\nu_\mu e}}{{\rm d}T_e}(E,T_e) \, X(E) \, {\rm d}E \tag{2.11}$$

The function $X^{\rm ES}_{\nu_\mu e}(T_e)$ is plotted in Fig. 1.

The relations (2.6) and (2.8) are based only on the assumption that solar $\nu_e$'s transform only into $\nu_\mu$ and/or $\nu_\tau$, i.e. $\sum_{\ell=e,\mu,\tau} {\rm P}_{\nu_e\to\nu_\ell}(E) = 1$. If there are also transitions of $\nu_e$'s into sterile states the sum $\sum_{\ell=e,\mu,\tau} {\rm P}_{\nu_e\to\nu_\ell}(E)$ is less than 1. In the next section we will consider in detail this case.

# 3 Sterile neutrinos

Let us consider now the general case of transitions of solar $\nu_e$'s into left-handed active and sterile neutrinos. In this case we have

$$\sum_{\ell=e,\mu,\tau} {\rm P}_{\nu_e\to\nu_\ell}(E) = 1 - {\rm P}_{\nu_e\to\nu_S}(E) \tag{3.1}$$

where $P_{\nu_e\to\nu_S}(E)$ is the probability of transition of $\nu_e$'s into all possible sterile states. Using Eq.(3.1) we obtain

$$\frac{\Sigma^{\mathrm{ES;CC}}}{X^{\mathrm{ES}}_{\nu_\mu e}} - \frac{N^{\mathrm{NC}}}{X^{\mathrm{NC}}_{\nu d}} = \frac{\int_{E^{\mathrm{NC}}_{\mathrm{th}}} \sigma^{\mathrm{NC}}_{\nu d}(E)\,\phi_{\nu_S}(E)\,\mathrm{d}E}{X^{\mathrm{NC}}_{\nu d}} - \frac{\int_{E^{\mathrm{ES}}_{\mathrm{th}}} \sigma_{\nu_\mu e}(E)\,\phi_{\nu_S}(E)\,\mathrm{d}E}{X^{\mathrm{ES}}_{\nu_\mu e}} \tag{3.2}$$

where

$$\Sigma^{\mathrm{ES;CC}} \equiv N^{\mathrm{ES}} - \int_{E^{\mathrm{ES}}_{\mathrm{th}}} \left[\sigma_{\nu_e e}(E) - \sigma_{\nu_\mu e}(E)\right] \phi_{\nu_e}(E)\,\mathrm{d}E \tag{3.3}$$

It is clear that in the case $P_{\nu_e\to\nu_S}(E) = 0$ the relation (3.2) coincides with Eq.(2.6). The left-hand side of Eq.(3.2) contains only quantities measurable by the observation of the CC, NC and ES processes. If it will occur that the left-hand side of this equation is not equal to zero, it will mean that solar $\nu_e$'s transform into sterile states. Let us stress that this test does not require any assumption on the initial flux of $^8$B $\nu_e$'s.

It is easy to generalize the relation (2.8) for the case of transitions of solar $\nu_e$'s into active as well as into sterile states. We have

$$\frac{\Sigma^{\mathrm{ES;CC}}(T_e)}{X^{\mathrm{ES}}_{\nu_\mu e}(T_e)} - \frac{N^{\mathrm{NC}}}{X^{\mathrm{NC}}_{\nu d}} = \frac{\int_{E^{\mathrm{NC}}_{\mathrm{th}}} \sigma^{\mathrm{NC}}_{\nu d}(E)\,\phi_{\nu_S}(E)\,\mathrm{d}E}{X^{\mathrm{NC}}_{\nu d}} - \frac{\int_{E_{\mathrm{m}}(T_e)} \frac{\mathrm{d}\sigma_{\nu_\mu e}}{\mathrm{d}T_e}(E,T_e)\,\phi_{\nu_S}(E)\,\mathrm{d}E}{X^{\mathrm{ES}}_{\nu_\mu e}(T_e)} \tag{3.4}$$

where

$$\Sigma^{\mathrm{ES;CC}}(T_e) \equiv n^{\mathrm{ES}}(T_e) - \int_{E_{\mathrm{m}}(T_e)} \left[\frac{\mathrm{d}\sigma_{\nu_e e}}{\mathrm{d}T_e}(E,T_e) - \frac{\mathrm{d}\sigma_{\nu_\mu e}}{\mathrm{d}T_e}(E,T_e)\right] \phi_{\nu_e}(E)\,\mathrm{d}E \tag{3.5}$$

and $X^{\mathrm{ES}}_{\nu_\mu e}(T_e)$ is given in Eq.(2.11).

Three comments are in order:

(1) The right-hand side of Eqs.(3.2) and (3.4) could be different from zero only if the probability $P_{\nu_e\to\nu_S}(E)$ depends on the neutrino energy $E$.

(2) According to our model calculation the two terms in the right-hand side of Eq.(3.2) tend to cancel each other. This cancellation depends on the value of the threshold energy $E^{\mathrm{ES}}_{\mathrm{th}}$. Higher values of $E^{\mathrm{ES}}_{\mathrm{th}}$ are preferable for testing the presence of sterile neutrinos in the solar neutrino flux.

(3) In order to estimate the size of the possible effect of sterile neutrinos, we have calculated the right-hand side of Eq.(3.4) (divided by the initial $^8$B $\nu_e$ flux $\Phi^{^8\mathrm{B}}_{\nu_e}$) in a model with $\nu_e$–$\nu_S$ mixing. The parameters of this model, obtained from the analysis of the existing data, are listed in Table 1. The results of our calculation are presented in Fig. 2: the two curves correspond to vacuum oscillations (see ref.[19]) and small mixing angle MSW transitions (see ref.[20]). Notice that the result depend on $T_e$ and for large $T_e$ the two terms in the right-hand side of Eq.(3.4) do not cancel each other.

Now we will obtain a number of inequalities of different nature which will allow to answer the question whether there are transitions of solar $\nu_e$'s into sterile states. We will also obtain lower bounds for the probability of $\nu_e \to \nu_S$ transitions.

In order to proof that there are sterile neutrinos in the solar neutrino flux on the earth it is sufficient to proof that the quantity $\sum_{\ell=e,\mu,\tau} \mathrm{P}_{\nu_e\to\nu_\ell}(E)$ or any average of this quantity is less than one. Consider the NC event rate. We have

$$\begin{aligned} N^{\mathrm{NC}} &= \Phi_{\nu_e}^{^8\mathrm{B}} \int_{E_{\mathrm{th}}^{\mathrm{NC}}} \sigma_{\nu d}^{\mathrm{NC}}(E)\, X(E) \sum_{\ell=e,\mu,\tau} \mathrm{P}_{\nu_e\to\nu_\ell}(E)\, \mathrm{d}E \\ &= \Phi_{\nu_e}^{^8\mathrm{B}}\, X_{\nu d}^{\mathrm{NC}} \left\langle \sum_{\ell=e,\mu,\tau} \mathrm{P}_{\nu_e\to\nu_\ell} \right\rangle_{\mathrm{NC}} \end{aligned} \tag{3.6}$$

In the case under consideration we cannot determine the value of the flux $\Phi_{\nu_e}^{^8\mathrm{B}}$ directly from the experimental data. But we can obtain a *lower bound* for this flux. In fact we have

$$\Phi_{\nu_e}^{^8\mathrm{B}} = \frac{\phi_{\nu_e}(E)}{X(E)\, \mathrm{P}_{\nu_e\to\nu_e}(E)} \tag{3.7}$$

From this relation it follows that

$$\Phi_{\nu_e}^{^8\mathrm{B}} \geq \left( \frac{\phi_{\nu_e}}{X} \right)_{\mathrm{max}} \tag{3.8}$$

where the subscript max indicates the maximum value in the explored energy range.

From Eqs.(3.6) and (3.8) we obtain

$$\left\langle \sum_{\ell=e,\mu,\tau} \mathrm{P}_{\nu_e\to\nu_\ell} \right\rangle_{\mathrm{NC}} \leq \frac{N^{\mathrm{NC}}}{X_{\nu d}^{\mathrm{NC}}\ (\phi_{\nu_e}/X)_{\mathrm{max}}} \tag{3.9}$$

Therefore, if the inequality

$$\frac{N^{\mathrm{NC}}}{X_{\nu d}^{\mathrm{NC}}\ (\phi_{\nu_e}/X)_{\mathrm{max}}} < 1 \tag{3.10}$$

will be satisfied, it will mean that solar $\nu_e$'s transform into sterile states [4]. In this case, for the average probability of the transition $\nu_e \to \nu_S$ we have

$$\langle \mathrm{P}_{\nu_e\to\nu_S} \rangle_{\mathrm{NC}} \geq 1 - \frac{N^{\mathrm{NC}}}{X_{\nu d}^{\mathrm{NC}}\ (\phi_{\nu_e}/X)_{\mathrm{max}}} \tag{3.11}$$

Let us discuss now the possibility to reveal the presence of sterile neutrinos in the solar neutrino flux on the earth from the measurement of the total ES event rate $N^{\mathrm{ES}}$ and the differential CC event rate $n^{\mathrm{CC}}(T_e)$. For the value of $\Sigma^{\mathrm{ES;CC}}$, defined in Eq.(3.3), we have

$$\begin{aligned}\Sigma^{\mathrm{ES;CC}} &= \Phi_{\nu_e}^{^8\mathrm{B}} \int_{E_{\mathrm{th}}^{\mathrm{ES}}} \sigma_{\nu_\mu e}(E)\, X(E) \sum_{\ell=e,\mu,\tau} \mathrm{P}_{\nu_e\to\nu_\ell}(E)\, \mathrm{d}E \\ &= \Phi_{\nu_e}^{^8\mathrm{B}}\, X_{\nu_\mu e}^{\mathrm{ES}} \left\langle \sum_{\ell=e,\mu,\tau} \mathrm{P}_{\nu_e\to\nu_\ell} \right\rangle_{\mathrm{ES}}\end{aligned} \tag{3.12}$$

[4] Obviously, in the case $N^{\mathrm{NC}}/[X_{\nu d}^{\mathrm{NC}}\ (\phi_{\nu_e}/X)_{\mathrm{max}}] \geq 1$ no conclusion on the existence of sterile neutrinos can be reached.

From Eqs.(3.12) and (3.8) it follows that

$$\left\langle \sum_{\ell=e,\mu,\tau} \mathrm{P}_{\nu_e\to\nu_\ell} \right\rangle_{\mathrm{ES}} \leq \frac{\Sigma^{\mathrm{ES;CC}}}{X_{\nu_\mu e}^{\mathrm{ES}}\, (\phi_{\nu_e}/X)_{\mathrm{max}}} \tag{3.13}$$

Thus, if the inequality

$$\frac{\Sigma^{\mathrm{ES;CC}}}{X_{\nu_\mu e}^{\mathrm{ES}}\, (\phi_{\nu_e}/X)_{\mathrm{max}}} < 1 \tag{3.14}$$

will be satisfied it will mean that there are sterile neutrinos in the solar neutrino flux on the earth. For the average value of the probability of the transition $\nu_e \to \nu_{\mathrm{S}}$ we have the following lower bound

$$\langle \mathrm{P}_{\nu_e\to\nu_{\mathrm{S}}} \rangle_{\mathrm{ES}} \geq 1 - \frac{\Sigma^{\mathrm{ES;CC}}}{X_{\nu_\mu e}^{\mathrm{ES}}\, (\phi_{\nu_e}/X)_{\mathrm{max}}} \tag{3.15}$$

Now we will obtain a *differential inequality* analogous to Eqs.(3.11) and (3.15). We have

$$\begin{aligned}\Sigma^{\mathrm{ES;CC}}(T_e) &= \Phi_{\nu_e}^{^8\mathrm{B}} \int_{E_{\mathrm{m}}(T_e)} \frac{\mathrm{d}\sigma_{\nu_\mu e}}{\mathrm{d}T_e}(E,T_e)\, X(E) \sum_{\ell=e,\mu,\tau} \mathrm{P}_{\nu_e\to\nu_\ell}(E)\, \mathrm{d}E \\ &= \Phi_{\nu_e}^{^8\mathrm{B}}\, X_{\nu_\mu e}^{\mathrm{ES}}(T_e) \left\langle \sum_{\ell=e,\mu,\tau} \mathrm{P}_{\nu_e\to\nu_\ell} \right\rangle_{\mathrm{ES};T_e}\end{aligned} \tag{3.16}$$

Here $X_{\nu_\mu e}^{\mathrm{ES}}(T_e)$ is given by Eq.(2.11) and the quantity $\Sigma^{\mathrm{ES;CC}}(T_e)$ is connected with the $\nu_e$ flux on the earth $\phi_{\nu_e}(E)$ and the spectrum of the recoil electrons in the ES process $n^{\mathrm{ES}}(T_e)$ by the relation (3.5). From Eqs.(3.12) and (3.8) we obtain

$$\left\langle \sum_{\ell=e,\mu,\tau} \mathrm{P}_{\nu_e\to\nu_\ell} \right\rangle_{\mathrm{ES};T_e} \leq \frac{\Sigma^{\mathrm{ES;CC}}(T_e)}{X_{\nu_\mu e}^{\mathrm{ES}}(T_e)\, (\phi_{\nu_e}/X)_{\mathrm{max}}} \tag{3.17}$$

Thus, if for some values of $T_e$ the inequality

$$\frac{\Sigma^{\mathrm{ES;CC}}(T_e)}{X_{\nu_\mu e}^{\mathrm{ES}}(T_e)\, (\phi_{\nu_e}/X)_{\mathrm{max}}} < 1 \tag{3.18}$$

will be satisfied, it will mean that sterile neutrinos exist. The lower bound of the average probability of the transition $\nu_e \to \nu_S$ depends on $T_e$ and is given by

$$\langle \mathrm{P}_{\nu_e \to \nu_S} \rangle_{\mathrm{ES};T_e} \geq 1 - \frac{\Sigma^{\mathrm{ES;CC}}(T_e)}{X^{\mathrm{ES}}_{\nu_\mu e}(T_e)\,(\phi_{\nu_e}/X)_{\mathrm{max}}} \tag{3.19}$$

We have calculated this lower bound and the quantity $\langle \mathrm{P}_{\nu_e \to \nu_S} \rangle_{\mathrm{ES};T_e}$ using a model with $\nu_e$–$\nu_S$ mixing. The results of our calculation are presented in Fig. 3: the two curves correspond to vacuum oscillations (see ref.[19]) and small mixing angle MSW transitions (see ref.[20]).

Let us stress that all the derived criterions for the existence of sterile neutrinos depend only on measurable quantities. If it will occur that at least one of these inequalities is satisfied, it will be *a model independent proof* of transitions of solar $\nu_e$'s into sterile states.

It is easy to see the physical meaning of the ratios that enter into the inequalities (3.10), (3.14) and (3.18). We have

$$\mathrm{P}^{\mathrm{max}}_{\nu_e \to \nu_e} = \frac{1}{\Phi^{^8\mathrm{B}}_{\nu_e}} \left( \frac{\phi_{\nu_e}}{X} \right)_{\mathrm{max}} \tag{3.20}$$

From Eq.(3.6), (3.12), (3.16) and (3.20) it follows that the quantities that enter into the inequalities (3.10), (3.14) and (3.18) coincide with

$$\frac{\left\langle \sum\limits_{\ell=e,\mu,\tau} \mathrm{P}_{\nu_e \to \nu_\ell} \right\rangle_{\mathrm{A}}}{\mathrm{P}^{\mathrm{max}}_{\nu_e \to \nu_e}} \tag{3.21}$$

Here $A = \mathrm{NC}, \ldots$. It is obvious that

$$\left\langle \sum_{\ell=e,\mu,\tau} \mathrm{P}_{\nu_e \to \nu_\ell} \right\rangle_{\mathrm{A}} \leq \frac{\left\langle \sum\limits_{\ell=e,\mu,\tau} \mathrm{P}_{\nu_e \to \nu_\ell} \right\rangle_{\mathrm{A}}}{\mathrm{P}^{\mathrm{max}}_{\nu_e \to \nu_e}} \tag{3.22}$$

Thus, if at least one of the inequalities (3.10), (3.14) and (3.18) will be satisfied, it will mean that the corresponding value of $\left\langle \sum\limits_{\ell=e,\mu,\tau} \mathrm{P}_{\nu_e \to \nu_\ell} \right\rangle_{\mathrm{A}}$ is less than one and solar neutrinos transform into sterile states.

Three comments are in order:

(1) The closer $\mathrm{P}^{\mathrm{max}}_{\nu_e \to \nu_e}$ is to one, the better is the test of sterility.

(2) Each probability in the right hand side of Eq.(3.22) is proportional to $1/\Phi^{^8\mathrm{B}}_{\nu_e}$. Thus, the ratio does not depend on the initial $^8$B $\nu_e$ flux. This is the basis of the model independent tests discussed here.

(3) If $\sum_{\ell=e,\mu,\tau} \mathrm{P}_{\nu_e\to\nu_\ell}$ does not depend on energy we have $\left\langle \sum_{\ell=e,\mu,\tau} \mathrm{P}_{\nu_e\to\nu_\ell} \right\rangle_{\mathrm{A}} \Big/ \mathrm{P}^{\max}_{\nu_e\to\nu_e} \geq 1$.
No conclusion about the existence of sterile neutrinos can be reached in this case.

In all the inequalities that we have derived we used the inequality (3.8) for the lower bound of the $^8$B $\nu_e$ flux (which is based on the inequality $\mathrm{P}_{\nu_e\to\nu_e}(E) \leq 1$). We can obtain other expressions for the lower bound of the $^8$B $\nu_e$ flux. In fact, from Eq.(3.6) we obtain [5]

$$\Phi^{^8\mathrm{B}}_{\nu_e} \geq \frac{N^{\mathrm{NC}}}{X^{\mathrm{NC}}_{\nu d}} \tag{3.23}$$

Furthermore, from Eq.(3.16) we obtain a third lower bound for the same quantity $\Phi^{^8\mathrm{B}}_{\nu_e}$:

$$\Phi^{^8\mathrm{B}}_{\nu_e} \geq \left( \frac{\Sigma^{\mathrm{ES;CC}}(T_e)}{X^{\mathrm{ES}}_{\nu_\mu e}(T_e)} \right)_{\max} \tag{3.24}$$

Using Eqs.(3.23) and (3.24) it is easy to obtain other inequalities whose experimental test will allow to get informations on the transitions of solar $\nu_e$'s into sterile states.

We will present one of them. From Eqs.(3.16) and (3.24) we have

$$\left\langle \sum_{\ell=e,\mu,\tau} \mathrm{P}_{\nu_e\to\nu_\ell} \right\rangle_{\mathrm{ES};T_e} \leq \frac{\Sigma^{\mathrm{ES;CC}}(T_e)}{X^{\mathrm{ES}}_{\nu_\mu e}(T_e) \left( \Sigma^{\mathrm{ES;CC}}(T'_e)/X^{\mathrm{ES}}_{\nu_\mu e}(T'_e) \right)_{\max}} \tag{3.25}$$

The right-hand side of the inequality (3.25) is not larger than one. It is clear that if the measurable quantity $\Sigma^{\mathrm{ES;CC}}(T_e)/X^{\mathrm{ES}}_{\nu_\mu e}(T_e)$ *depends on energy* it would mean that solar $\nu_e$'s transform into sterile states. From our point of view this is one of the simplest tests of sterility.

For the lower bound of the probability of the transition $\nu_e \to \nu_S$ we obtain

$$\langle \mathrm{P}_{\nu_e\to\nu_S} \rangle_{\mathrm{ES};T_e} \geq 1 - \frac{\Sigma^{\mathrm{ES;CC}}(T_e)}{X^{\mathrm{ES}}_{\nu_\mu e}(T_e) \left( \Sigma^{\mathrm{ES;CC}}(T'_e)/X^{\mathrm{ES}}_{\nu_\mu e}(T'_e) \right)_{\max}} \tag{3.26}$$

In Super-Kamiokande and other future solar neutrino experiments (ICARUS, etc.) a large number of solar neutrino induced ES events will be observed. If from these data the differential ES event rate $n^{\mathrm{ES}}(E)$ will be determined as a function of the neutrino energy $E$, new possibilities for testing the existence of sterile neutrinos will emerge. In fact, a

[5] Let us notice that this inequality corresponds to the equality (2.2) which is valid in the case of transitions of solar $\nu_e$'s into active $\nu_\mu$ and/or $\nu_\tau$. Instead of the relation (2.4) for the survival probability we have the following inequality

$$\mathrm{P}_{\nu_e\to\nu_e}(E) \leq \alpha(E)\,\frac{\phi_{\nu_e}(E)}{N^{\mathrm{NC}}}$$

where $\alpha(E)$ is given by Eq.(2.5).

measurement of $n^{\rm ES}(E)$ and $n^{\rm CC}(T_e)$ will allow to determine the differential flux of all types of active neutrinos on the earth:

$$\sum_{\ell=e,\mu,\tau} \phi_{\nu_\ell}(E) = \frac{n^{\rm ES}(E)}{\sigma_{\nu_\mu e}(E)} + \left(1 - \frac{\sigma_{\nu_e e}(E)}{\sigma_{\nu_\mu e}(E)}\right) \phi_{\nu_e}(E) \tag{3.27}$$

From Eqs.(2.1) and (3.1) we obtain

$$\frac{1}{X(E)} \sum_{\ell=e,\mu,\tau} \phi_{\nu_\ell}(E) = \Phi^{^8{\rm B}}_{\nu_e} \left[1 - {\rm P}_{\nu_e\to\nu_S}(E)\right] \tag{3.28}$$

If it will occur that the left-hand side of this equation, which contains only measurable quantities, depends on energy, then it will mean that ${\rm P}_{\nu_e\to\nu_S}(E) \neq 0$, i.e. there are sterile neutrinos in the flux of solar neutrinos on the earth.

From Eq.(3.28) we derive the following lower bound for the $^8$B $\nu_e$ flux:

$$\Phi^{^8{\rm B}}_{\nu_e} \geq \left( \sum_{\ell=e,\mu,\tau} \phi_{\nu_\ell} \Big/ X \right)_{\rm max} \tag{3.29}$$

Furthermore, for the probability of the transition $\nu_e \to \nu_S$ as a function of the neutrino energy E we find the lower bound

$${\rm P}_{\nu_e\to\nu_S}(E) \geq 1 - \frac{\sum\limits_{\ell=e,\mu,\tau} \phi_{\nu_\ell}(E)}{X(E) \left( \sum\limits_{\ell=e,\mu,\tau} \phi_{\nu_\ell} \Big/ X \right)_{\rm max}} \tag{3.30}$$

In conclusion let us notice that the lower bounds (3.8), (3.23), (3.24) and (3.29) could allow us to test the SSM in a model independent way. In fact, if it will occur that the flux $\Phi^{^8{\rm B}}_{\nu_e}({\rm SSM})$ predicted by the SSM is less than any of the lower bounds, it will mean that the solar neutrino flux predicted by the SSM is lower than the real flux [6].

# 4 Neutrino Spin-Flavor Precession

In this section we will discuss possibilities of a model independent treatment of the data from future solar neutrino experiments in the case of spin and/or spin-flavor precession of solar $\nu_e$'s due to anomalously large neutrino magnetic moments. The Hamiltonian of interaction of the electromagnetic field with Dirac neutrinos have the following general form:

$$\mathcal{H}^{\rm D}_I = - \sum_{\ell',\ell} \mu^{\rm D}_{\ell'\ell} \, \bar\nu_{\ell' R} \, \sigma^{\alpha\beta} \, \nu_{\ell L} \, F_{\alpha\beta} + {\rm h.c.} \tag{4.1}$$

[6] This possibility seems rather unusual. Notice however that it could be realized in the case of a strong energy dependence of the survival probability.

where $F_{\alpha\beta}$ is the strength tensor of the electromagnetic field and $\mu^{\rm D}$ is the (non-diagonal) matrix of magnetic moments. Due to the interaction (4.1) solar $\nu_e$'s can be transformed into sterile right-handed $\nu_{\ell R}$ ($\ell = e, \mu, \tau$), quanta of right-handed fields.

In the Majorana case the Hamiltonian of interaction with the electromagnetic field have the following general form:

$$\mathcal{H}_I^{\rm M} = - \sum_{\ell',\ell} \mu_{\ell'\ell}^{\rm M} \, \overline{(\nu_{\ell' L})^c} \, \sigma^{\alpha\beta} \, \nu_{\ell L} \, F_{\alpha\beta} + \text{h.c.} \tag{4.2}$$

where $\mu_{\ell'\ell}^{\rm M} = -\mu_{\ell\ell'}^{\rm M}$ and $(\nu_{\ell' L})^c = \mathcal{C} \overline{\nu_{\ell' L}}^T$ is the charge-conjugated field. Due to the interaction (4.2) solar $\nu_e$'s can be transformed into active right-handed antineutrinos $\bar{\nu}_\ell$.

If it will occur that the CC event rate depends on time periodically, it will be an evidence that neutrinos have large magnetic moments. We will discuss here which additional informations about magnetic moments can be extracted from a measurement of the NC and ES event rates. We will show that it will be possible to distinguish Dirac from Majorana magnetic moments.

Let us notice that in SNO it is planned [21] to search for $\bar{\nu}_e$ from the sun with the observation of the reaction

$$\bar{\nu}_e + d \to e^+ + n + n \tag{4.3}$$

Direct transitions $\nu_e \to \bar{\nu}_e$ are forbidden by CPT invariance ($\mu_{\ell\ell}^{\rm M} = 0$). However, sizable $\nu_e \to \bar{\nu}_e$ transitions can occur under special conditions if both spin-flavor precession and the MSW or vacuum oscillations mechanisms are operating [22] [7].

## 4.1 Neutral Current

Let us consider first the NC process (1.2). In the case of Majorana neutrino magnetic moments the integral NC event rate is given by

$$N^{\rm NC} = \int_{E_{\rm th}^{\rm NC}} \sigma_{\nu d}^{\rm NC}(E) \sum_{\ell=e,\mu,\tau} \phi_{\nu_\ell}(E) \, \mathrm{d}E + \int_{E_{\rm th}^{\rm NC}} \sigma_{\bar{\nu} d}^{\rm NC}(E) \sum_{\ell=e,\mu,\tau} \phi_{\bar{\nu}_\ell}(E) \, \mathrm{d}E \tag{4.4}$$

where $\sigma_{\nu d}^{\rm NC}(E)$ and $\sigma_{\bar{\nu} d}^{\rm NC}(E)$ are the cross sections of the processes $\nu d \to \nu p n$ and $\bar{\nu} d \to \bar{\nu} p n$, $\sum_{\ell=e,\mu,\tau} \phi_{\nu_\ell}(E)$ and $\sum_{\ell=e,\mu,\tau} \phi_{\bar{\nu}_\ell}(E)$ are the fluxes of all types of neutrinos and antineutrinos on the earth. Taking into account that

$$\sum_{\ell=e,\mu,\tau} \phi_{\nu_\ell}(E) + \sum_{\ell=e,\mu,\tau} \phi_{\bar{\nu}_\ell}(E) = \phi_{\nu_e}^{^8\rm B}(E) \tag{4.5}$$

[7] The Mont Blanc collaboration has obtained the following upper bound for the flux of $\bar{\nu}_e$ in the energy range $9\,\text{MeV} \leq E \leq 20\,\text{MeV}$: $\Phi_{\bar{\nu}_e} < 8.2 \times 10^4\,\text{cm}^{-2}\,\text{sec}^{-1}$ [23]. From the analysis of the background in the Kamiokande experiment the following upper bound for the flux of high-energy $\bar{\nu}_e$ with $E \geq 10.6\,\text{MeV}$ was obtained: $\Phi_{\bar{\nu}_e} < 6.1 \times 10^4\,\text{cm}^{-2}\,\text{sec}^{-1}$ [24].

from Eq.(2.1) and Eq.(4.4) we obtain

$$\frac{N^{\rm NC}}{\Phi_{\nu_e}^{^8\rm B}\,\overline{X}^{\rm NC}} = 1 + I^{\rm NC} \tag{4.6}$$

where

$$I^{\rm NC} \equiv \frac{1}{\overline{X}^{\rm NC}} \int_{E_{\rm th}^{\rm NC}} \left[ \frac{\sigma_{\nu d}^{\rm NC}(E) - \sigma_{\bar\nu d}^{\rm NC}(E)}{2} \right] X(E) \sum_{\ell=e,\mu,\tau} [{\rm P}_{\nu_e\to\nu_\ell}(E) - {\rm P}_{\nu_e\to\bar\nu_\ell}(E)]\, {\rm d}E \tag{4.7}$$

Here $\sum_{\ell=e,\mu,\tau} {\rm P}_{\nu_e\to\nu_\ell}(E)$ $\left( \sum_{\ell=e,\mu,\tau} {\rm P}_{\nu_e\to\bar\nu_\ell}(E) \right)$ is the transition probability of solar $\nu_e$'s into neutrinos (antineutrinos) of all types and

$$\overline{X}^{\rm NC} \equiv \frac{X_{\nu d}^{\rm NC} + X_{\bar\nu d}^{\rm NC}}{2} \tag{4.8}$$

with

$$X_{\nu d(\bar\nu d)}^{\rm NC} \equiv \int_{E_{\rm th}^{\rm NC}} \sigma_{\nu d(\bar\nu d)}^{\rm NC}(E)\, X(E)\, {\rm d}E \tag{4.9}$$

The cross sections of the processes $\nu d \to \nu n p$ and $\bar\nu d \to \bar\nu n p$ where calculated by several groups and reviewed in ref.[17]. Using the results presented in ref.[17], we obtained

$$X_{\nu d}^{\rm NC} = 4.72 \times 10^{-43}\,{\rm cm}^2 \tag{4.10}$$

$$X_{\bar\nu d}^{\rm NC} = 4.51 \times 10^{-43}\,{\rm cm}^2 \tag{4.11}$$

It is easy to see that the value of the integral $I^{\rm NC}$ is very small. In fact, for the absolute value of this integral we have the following upper bound:

$$\left| I^{\rm NC} \right| \le \frac{X_{\nu d}^{\rm NC} - X_{\bar\nu d}^{\rm NC}}{X_{\nu d}^{\rm NC} + X_{\bar\nu d}^{\rm NC}} \simeq 2 \times 10^{-2} \tag{4.12}$$

The upper bound of the integral $I^{\rm NC}$ is so small because the cross sections of the processes $\nu d \to \nu n p$ and $\bar\nu d \to \bar\nu n p$ are very close to each other in the energy region near the threshold. The argument in favor of this fact is rather general: from symmetry considerations it follows that near the threshold (if only the s-state of the final nucleons is taken into account) the vector current does not contribute to the matrix elements of the processes $\nu d \to \nu n p$ and $\bar\nu d \to \bar\nu n p$ and the cross sections $\sigma_{\nu d}^{\rm NC}$ and $\sigma_{\bar\nu d}^{\rm NC}$ are equal. The corrections due to higher states are small in the relevant energy region (see ref.[17] and references therein). Thus, the term $I^{\rm NC}$ in Eq.(4.6) can be safely neglected and, in the case of Majorana magnetic moments, we come to the following conclusions:

(1) The NC event rate does not depend on time (within less than 2%).

(2) The flux of the initial $^8$B $\nu_e$'s is given by

$$\Phi_{\nu_e}^{^8\mathrm{B}} \simeq \frac{N^{\mathrm{NC}}}{\overline{X}^{\mathrm{NC}}} \tag{4.13}$$

and therefore can be determined directly from the experimental data [8].

(3) It is possible to obtain the $\nu_e$ survival probability directly from measurable quantities:

$$\begin{aligned} \mathrm{P}_{\nu_e\to\nu_e}(E) &= 1 - \sum_{\ell=\mu,\tau} \mathrm{P}_{\nu_e\to\nu_\ell}(E) - \sum_{\ell=e,\mu,\tau} \mathrm{P}_{\nu_e\to\bar{\nu}_\ell}(E) \\ &= \frac{\phi_{\nu_e}(E)\,\overline{X}^{\mathrm{NC}}}{X(E)\,N^{\mathrm{NC}}} \end{aligned} \tag{4.14}$$

where $\phi_{\nu_e}(E)$ is the flux of $\nu_e$ on the earth, which can be determined from the CC event rate.

Consider now the case of Dirac neutrino magnetic moments. In this case

$$N^{\mathrm{NC}} = \int_{E_{\mathrm{th}}^{\mathrm{NC}}} \sigma_{\nu d}^{\mathrm{NC}}(E) \sum_{\ell=e,\mu,\tau} \phi_{\nu_\ell}(E)\,\mathrm{d}E \tag{4.15}$$

where $\sum\limits_{\ell=e,\mu,\tau} \phi_{\nu_\ell}(E) = \phi_{\nu_e}^{^8\mathrm{B}}(E) - \phi_{\nu_S}(E)$ and $\phi_{\nu_S}(E)$ is the flux of sterile right-handed neutrinos on the earth. It is clear that in the Dirac case $N^{\mathrm{NC}}$ will depend on time. Hence, a time dependence of the CC *and* NC event rates will be a signal that neutrinos have large Dirac magnetic moments.

For the average transition probability of solar $\nu_e$'s into sterile states we obtain the following model independent lower bound:

$$\begin{aligned} \left\langle \sum_{\ell=e,\mu,\tau} \mathrm{P}_{\nu_e\to\nu_{\ell R}} \right\rangle_{\mathrm{NC}} &\equiv \frac{1}{X_{\nu d}^{\mathrm{NC}}} \int_{E_{\mathrm{th}}^{\mathrm{NC}}} \sigma_{\nu d}^{\mathrm{NC}}(E)\,X(E) \sum_{\ell=e,\mu,\tau} \mathrm{P}_{\nu_e\to\nu_{\ell R}}(E)\,\mathrm{d}E \\ &\geq 1 - \frac{N^{\mathrm{NC}}}{X_{\nu d}^{\mathrm{NC}}\,(\phi_{\nu_e}/X)_{\max}} \end{aligned} \tag{4.16}$$

Let us stress that, unlike the case of Eq.(3.11), in the case of spin and/or spin-flavor transitions due to Dirac magnetic moments the lower bound (4.16) will depend on time.

## 4.2 Elastic Scattering

Let us consider now the ES process (1.3). Using Eq.(2.1), in the case of Majorana neutrinos we have

$$\frac{\tilde{\Sigma}^{\mathrm{ES;CC}}}{\Phi_{\nu_e}^{^8\mathrm{B}}\,\overline{X}^{\mathrm{ES}}} = 1 + I^{\mathrm{ES}} \tag{4.17}$$

[8]Notice that the expression (4.13) for $\Phi_{\nu_e}^{^8\mathrm{B}}$ coincides with Eq.(2.2) obtained for the case of transitions of solar $\nu_e$'s only into $\nu_\mu$ and/or $\nu_\tau$ due to usual neutrino mixing.

where

$$
\begin{aligned}
\widetilde{\Sigma}^{\mathrm{ES;CC}} \equiv N^{\mathrm{ES}} & - \int_{E_{\mathrm{th}}^{\mathrm{ES}}} \left[\sigma_{\nu_e e}(E) - \sigma_{\nu_\mu e}(E)\right] \phi_{\nu_e}(E)\, \mathrm{d}E \\
& - \int_{E_{\mathrm{th}}^{\mathrm{ES}}} \left[\sigma_{\bar{\nu}_e e}(E) - \sigma_{\bar{\nu}_\mu e}(E)\right] \phi_{\bar{\nu}_e}(E)\, \mathrm{d}E
\end{aligned} \tag{4.18}
$$

$N^{\mathrm{ES}}$ is the integral ES event rate, $\sigma_{\nu_\ell e}(E)$ ($\sigma_{\bar{\nu}_\ell e}(E)$) is the total cross section of the process $\nu_\ell e \to \nu_\ell e$ ($\bar{\nu}_\ell e \to \bar{\nu}_\ell e$) with $\ell = e, \mu$, and

$$
I^{\mathrm{ES}} \equiv \frac{1}{\overline{X}^{\mathrm{ES}}} \int_{E_{\mathrm{th}}^{\mathrm{ES}}} \left[\frac{\sigma_{\nu_\mu e}(E) - \sigma_{\bar{\nu}_\mu e}(E)}{2}\right] X(E) \sum_{\ell=e,\mu,\tau} \left[\mathrm{P}_{\nu_e \to \nu_\ell}(E) - \mathrm{P}_{\nu_e \to \bar{\nu}_\ell}(E)\right] \mathrm{d}E \tag{4.19}
$$

where

$$
\overline{X}^{\mathrm{ES}} \equiv \frac{X_{\nu_\mu e}^{\mathrm{ES}} + X_{\bar{\nu}_\mu e}^{\mathrm{ES}}}{2} \tag{4.20}
$$

with

$$
X_{\nu_\mu e(\bar{\nu}_\mu e)}^{\mathrm{ES}} \equiv \int_{E_{\mathrm{th}}^{\mathrm{ES}}} \sigma_{\nu_\mu e(\bar{\nu}_\mu e)}(E)\, X(E)\, \mathrm{d}E \tag{4.21}
$$

For $T_e^{\mathrm{th}} = 4.5\,\mathrm{MeV}$ we obtained

$$
X_{\nu_\mu e}^{\mathrm{ES}} = 3.23 \times 10^{-45}\,\mathrm{cm}^2 \tag{4.22}
$$

$$
X_{\bar{\nu}_\mu e}^{\mathrm{ES}} = 2.57 \times 10^{-45}\,\mathrm{cm}^2 \tag{4.23}
$$

Using these values, we have the following upper bound for the absolute value of the integral $I^{\mathrm{ES}}$:

$$
\left|I^{\mathrm{ES}}\right| \leq \frac{X_{\nu_\mu e}^{\mathrm{ES}} - X_{\bar{\nu}_\mu e}^{\mathrm{ES}}}{X_{\nu_\mu e}^{\mathrm{ES}} + X_{\bar{\nu}_\mu e}^{\mathrm{ES}}} \simeq 0.11 \tag{4.24}
$$

Let us notice that the real value of $\left|I^{\mathrm{ES}}\right|$ can be significantly smaller than the upper bound given in Eq.(4.24). We calculated the integral $I^{\mathrm{ES}}$ in the simplest model with two non-mixed massive Majorana neutrinos and a large transition magnetic moment $\mu_{e\mu}$. In ref.[5] it was shown that the existing solar neutrino data can be described by this model with $\mu_{e\mu} \simeq 10^{-11}\,\mu_{\mathrm{B}}$ under specific assumptions for the magnetic field of the sun. For the average value of $\Delta m^2$ found in ref.[5] ($\Delta m^2 = 10^{-8}\,\mathrm{eV}^2$) we obtained that $-7.5 \times 10^{-2} \leq I^{\mathrm{ES}} \leq 8.7 \times 10^{-3}$, where the lower and upper bounds correspond to high and low solar activity, respectively.

The time dependence of the quantity $\widetilde{\Sigma}^{\mathrm{ES;CC}}$ is determined by the integral $I^{\mathrm{ES}}$, which is less than $\simeq 10\%$. Neglecting $I^{\mathrm{ES}}$ in Eq.(4.17), we obtain the following approximate expression for the initial flux of $^8\mathrm{B}$ neutrinos:

$$
\Phi_{\nu_e}^{^8\mathrm{B}} \simeq \frac{\widetilde{\Sigma}^{\mathrm{ES;CC}}}{\overline{X}^{\mathrm{ES}}} \tag{4.25}
$$

Thus, in the case of Majorana magnetic moments, the initial flux of $^8$B neutrinos can be determined in two independent ways: from the NC event rate (see Eq.(4.13)) and from the ES and CC event rates (see Eq.(4.25)). Therefore, independently from the value of the initial neutrino flux, we have the following approximate relation between measurable quantities:

$$N^{\rm NC} \simeq \frac{\overline{X}^{\rm NC}}{\overline{X}^{\rm ES}} \, \widetilde{\Sigma}^{\rm ES;CC} \tag{4.26}$$

This relation is a generalization of the analogous relation (2.6) that was obtained for the case in which only active neutrinos $\nu_e$, $\nu_\mu$, $\nu_\tau$ are present in the flux of solar neutrinos on the earth (in that case the relation is exact).

Let us notice that in the case of Majorana magnetic moments the ES event rate will depend on time. This time dependence is determined by the time dependence of the CC event rate.

Let us consider now the case of Dirac neutrino magnetic moments. In this case we have

$$\frac{\Sigma^{\rm ES;CC}}{\Phi^{^8\rm B}_{\nu_e} \, X^{\rm ES}_{\nu_\mu e}} = 1 - \frac{1}{X^{\rm ES}_{\nu_\mu e}} \int_{E^{\rm ES}_{\rm th}} \sigma_{\nu_\mu e}(E) \, X(E) \sum_{\ell=e,\mu,\tau} {\rm P}_{\nu_e \to \nu_{\ell R}}(E) \, {\rm d}E \tag{4.27}$$

where

$$\Sigma^{\rm ES;CC} \equiv N^{\rm ES} - \int_{E^{\rm ES}_{\rm th}} \Big( \sigma_{\nu_e e}(E) - \sigma_{\nu_\mu e}(E) \Big) \, \phi_{\nu_e}(E) \, {\rm d}E \tag{4.28}$$

In the case under consideration the quantity $\Sigma^{\rm ES;CC}$ will depend on time. From Eq.(4.27), we obtain the following lower bound for the average transition probability of solar $\nu_e$'s into all possible right-handed sterile states:

$$\begin{aligned} \left\langle \sum_{\ell=e,\mu,\tau} {\rm P}_{\nu_e \to \nu_{\ell R}} \right\rangle_{\rm ES} &\equiv \frac{1}{X^{\rm ES}_{\nu_\mu e}} \int_{E^{\rm ES}_{\rm th}} \sigma_{\nu_\mu e}(E) \, X(E) \sum_{\ell=e,\mu,\tau} {\rm P}_{\nu_e \to \nu_{\ell R}}(E) \, {\rm d}E \\ &\geq 1 - \frac{\Sigma^{\rm ES;CC}}{X^{\rm ES}_{\nu_\mu e} \, (\phi_{\nu_e}/X)_{\rm max}} \end{aligned} \tag{4.29}$$

In the case of Dirac magnetic moments, instead of relation (4.26) we have

$$N^{\rm NC} = \frac{X^{\rm NC}_{\nu d}}{X^{\rm ES}_{\nu_\mu e}} \, \Sigma^{\rm ES;CC} + \beta \tag{4.30}$$

where

$$\beta \equiv \frac{X^{\rm NC}_{\nu d}}{X^{\rm ES}_{\nu_\mu e}} \int_{E^{\rm ES}_{\rm th}} \sigma_{\nu_\mu e}(E) \, \phi_{\nu_S}(E) \, {\rm d}E - \int_{E^{\rm NC}_{\rm th}} \sigma^{\rm NC}_{\nu d}(E) \, \phi_{\nu_S}(E) \, {\rm d}E \tag{4.31}$$

Let us assume that $\bar{\nu}_e$ is not observed. If the relation (4.26) is violated and $\beta$ depends on time, we will have an additional argument in favor of Dirac magnetic moments. Notice, however, that, according to our model calculations, the two terms in the right-hand side of Eq.(4.31) could cancel each other.

## 4.3 On the transition probability of $\nu_e$ into $\bar{\nu}_\mu$ and/or $\bar{\nu}_\tau$

Let us consider the case of Majorana magnetic moments (the CC event rate depends on time but the NC event rate is constant). As we have seen in section 4.1, from the data on the CC and NC reactions (and the process (4.3)) it will be possible to determine the sum of the probabilities of the transitions of initial $\nu_e$'s into $\nu_\mu$, $\nu_\tau$, $\bar{\nu}_\mu$, $\bar{\nu}_\tau$ (see Eq.(4.14)). In this section we will discuss possibilities to obtain from the SNO and S-K data an information about the transition probability $\sum_{\ell=\mu,\tau} \mathrm{P}_{\nu_e \to \bar{\nu}_\ell}(E)$. With the help of Eq.(4.13) we obtain

$$\frac{\Omega^{\mathrm{ES;CC}}}{N^{\mathrm{NC}}} \frac{\overline{X}^{\mathrm{NC}}}{X^{\mathrm{ES}}_{\nu_\mu e}} \simeq 1 - \widetilde{I}^{\mathrm{ES}} \tag{4.32}$$

where

$$\begin{aligned} \Omega^{\mathrm{ES;CC}} \equiv N^{\mathrm{ES}} &- \int_{E^{\mathrm{ES}}_{\mathrm{th}}} \left[\sigma_{\nu_e e}(E) - \sigma_{\nu_\mu e}(E)\right] \phi_{\nu_e}(E)\,\mathrm{d}E \\ &- \int_{E^{\mathrm{ES}}_{\mathrm{th}}} \left[\sigma_{\bar{\nu}_e e}(E) - \sigma_{\nu_\mu e}(E)\right] \phi_{\bar{\nu}_e}(E)\,\mathrm{d}E \end{aligned} \tag{4.33}$$

and

$$\widetilde{I}^{\mathrm{ES}} \equiv \frac{1}{X^{\mathrm{ES}}_{\nu_\mu e}} \int_{E^{\mathrm{ES}}_{\mathrm{th}}} \left[\sigma_{\nu_\mu e}(E) - \sigma_{\bar{\nu}_\mu e}(E)\right] X(E) \sum_{\ell=\mu,\tau} \mathrm{P}_{\nu_e \to \bar{\nu}_\ell}(E)\,\mathrm{d}E \tag{4.34}$$

It is easy to see that

$$0 \leq \widetilde{I}^{\mathrm{ES}} \leq \frac{X^{\mathrm{ES}}_{\nu_\mu e} - X^{\mathrm{ES}}_{\bar{\nu}_\mu e}}{X^{\mathrm{ES}}_{\nu_\mu e}} = 0.21 \tag{4.35}$$

Hence, in order to obtain an information about the transition probability $\sum_{\ell=\mu,\tau} \mathrm{P}_{\nu_e \to \bar{\nu}_\ell}(E)$, it is necessary to know the measurable quantity $\Omega^{\mathrm{ES;CC}}\overline{X}^{\mathrm{NC}}/N^{\mathrm{NC}}X^{\mathrm{ES}}_{\nu_\mu e}$ with an accuracy better than $\simeq$ 20%. If this accuracy will be achieved, the average probability of the transition of solar $\nu_e$'s into $\bar{\nu}_\mu$ and/or $\bar{\nu}_\tau$ can be determined directly from the experimental data:

$$\begin{aligned} \left\langle \sum_{\ell=\mu,\tau} \mathrm{P}_{\nu_e \to \bar{\nu}_\ell} \right\rangle_{\mathrm{ES}} &\equiv \frac{\int_{E^{\mathrm{ES}}_{\mathrm{th}}} \left[\sigma_{\nu_\mu e}(E) - \sigma_{\bar{\nu}_\mu e}(E)\right] X(E) \sum_{\ell=\mu,\tau} \mathrm{P}_{\nu_e \to \bar{\nu}_\ell}(E)\,\mathrm{d}E}{X^{\mathrm{ES}}_{\nu_\mu e} - X^{\mathrm{ES}}_{\bar{\nu}_\mu e}} \\ &\simeq \frac{X^{\mathrm{ES}}_{\nu_\mu e}}{X^{\mathrm{ES}}_{\nu_\mu e} - X^{\mathrm{ES}}_{\bar{\nu}_\mu e}} \left(1 - \frac{\Omega^{\mathrm{ES;CC}}}{N^{\mathrm{NC}}} \frac{\overline{X}^{\mathrm{NC}}}{X^{\mathrm{ES}}_{\nu_\mu e}}\right) \end{aligned} \tag{4.36}$$

It is easy to derive also a differential relation analogous to Eq.(4.36), which gives the average probability of the transition of solar $\nu_e$'s into $\bar{\nu}_\mu$ and/or $\bar{\nu}_\tau$ as a function of the kinetic energy $T_e$ of the recoil electrons:

$$\left\langle \sum_{\ell=\mu,\tau} \mathrm{P}_{\nu_e\to\bar{\nu}_\ell} \right\rangle_{\mathrm{ES};T_e} \equiv \frac{\displaystyle\int_{E_\mathrm{m}(T_e)} \left[ \frac{\mathrm{d}\sigma_{\nu_\mu e}}{\mathrm{d}T_e}(E,T_e) - \frac{\mathrm{d}\sigma_{\bar{\nu}_\mu e}}{\mathrm{d}T_e}(E,T_e) \right] X(E) \sum_{\ell=\mu,\tau} \mathrm{P}_{\nu_e\to\bar{\nu}_\ell}(E)\,\mathrm{d}E}{X^{\mathrm{ES}}_{\nu_\mu e}(T_e) - X^{\mathrm{ES}}_{\bar{\nu}_\mu e}(T_e)}$$

$$\simeq \frac{X^{\mathrm{ES}}_{\nu_\mu e}(T_e)}{X^{\mathrm{ES}}_{\nu_\mu e}(T_e) - X^{\mathrm{ES}}_{\bar{\nu}_\mu e}(T_e)} \left( 1 - \frac{\Omega^{\mathrm{ES;CC}}(T_e)}{N^{\mathrm{NC}}} \, \frac{\overline{X}^{\mathrm{NC}}}{X^{\mathrm{ES}}_{\nu_\mu e}(T_e)} \right) \tag{4.37}$$

where

$$\begin{aligned} \Omega^{\mathrm{ES;CC}}(T_e) \equiv n^{\mathrm{ES}}(T_e) &- \int_{E_\mathrm{m}(T_e)} \left[ \frac{\mathrm{d}\sigma_{\nu_e e}}{\mathrm{d}T_e}(E,T_e) - \frac{\mathrm{d}\sigma_{\nu_\mu e}}{\mathrm{d}T_e}(E,T_e) \right] \phi_{\nu_e}(E)\,\mathrm{d}E \\ &- \int_{E_\mathrm{m}(T_e)} \left[ \frac{\mathrm{d}\sigma_{\bar{\nu}_e e}}{\mathrm{d}T_e}(E,T_e) - \frac{\mathrm{d}\sigma_{\nu_\mu e}}{\mathrm{d}T_e}(E,T_e) \right] \phi_{\bar{\nu}_e}(E)\,\mathrm{d}E \end{aligned} \tag{4.38}$$

# 5 Conclusions

From our point of view the solar neutrino problem will be unambiguously solved when a complete set of solar neutrino measurements will be done. By a complete set we mean a set of measurement in which the total neutrino flux and the $\nu_e$ survival probability are measured *independently*. We have shown here that in the future SNO and S-K experiments, in which high energy solar neutrinos originating from $^8$B will be detected, a complete set of solar neutrino measurements could be performed.

The program of a complete set of solar neutrino measurements cannot be realized in principle if part of the solar neutrinos transform into sterile states. We have considered here the problem of sterile neutrinos in detail. We have derived a number of relations between measurable quantities the test of which could allow to reveal in a model independent way the presence of transitions of solar $\nu_e$'s into sterile states. Let us stress that a discovery of these transitions would be a discovery of an effect beyond the Standard Model.

We have considered here also possible spin-flavor precession of neutrinos in the magnetic field and in the matter of the sun. It is shown that future solar neutrino experiments will allow to distinguish between the cases of Dirac and Majorana magnetic moments.

In future experiments solar neutrinos will be detected through the observation of CC, NC as well as CC and NC (ES) reactions. These data will allow to reveal directly whether there are $\nu_\mu$ and/or $\nu_\tau$ in the flux of solar neutrinos on the earth. The ratio

$$R^{\mathrm{NC}} = 1 - \frac{\displaystyle\int_{E^{\mathrm{NC}}_{\mathrm{th}}} \sigma^{\mathrm{NC}}_{\nu d}(E)\,\phi_{\nu_e}(E)\,\mathrm{d}E}{N^{\mathrm{NC}}} \tag{5.1}$$

characterizes the relative contribution of $\nu_\mu$ and/or $\nu_\tau$ in the NC event rate. It is obvious that $R^{\rm NC} = 0$ if there are no $\nu_\mu$ and/or $\nu_\tau$ in the flux of solar neutrinos on the earth and $R^{\rm NC} = 1$ if there are only $\nu_\mu$ and/or $\nu_\tau$ (and maybe sterile neutrinos). The results of our calculation of $R^{\rm NC}$ in a model with mixing of two neutrino types ($\nu_e$ and $\nu_\mu$ or $\nu_\tau$) are depicted in Fig. 4. The two regions of values of $\Delta m^2$ and $\sin^2 2\vartheta$ which were obtained from the analysis of the existing data [4] are also plotted in Fig. 4. It can be seen from Fig. 4 that the value of the ratio $R^{\rm NC}$ can be as high as 0.6–0.8.

The ratio

$$r^{\rm ES}(T_e) = 1 - \frac{\displaystyle\int_{E_{\rm m}(T_e)} \frac{{\rm d}\sigma_{\nu_e e}}{{\rm d}T_e}(E,T_e)\,\phi_{\nu_e}(E)\,{\rm d}E}{n^{\rm ES}(T_e)} \tag{5.2}$$

characterizes the relative contribution of $\nu_\mu$ and/or $\nu_\tau$ to the differential ES event rate. The results of a calculation of $r^{\rm ES}(T_e)$ in a model with mixing of two neutrino types ($\nu_e$ and $\nu_\mu$ or $\nu_\tau$) are depicted in Fig. 5.

# Acknowledgments

It is a pleasure for us to express our gratitude to Vittorio de Alfaro and Wanda Alberico for very useful discussions. We would like to thank Naoya Hata for his useful comments. We thank the Institute for Nuclear Theory at the University of Washington for its hospitality and the Department of Energy for partial support during the completion of this work.

# A On the determination of the flux of $\nu_\mu$ and $\nu_\tau$

In section 3 we saw how important will be the knowledge of the flux on the earth of all neutrino types as a function of neutrino energy. In this appendix we present a possible method of restoration of the flux of $\nu_\mu$ and/or $\nu_\tau$ from the ES and CC data.

If the ES event rate is measured with a detector having a good energy and angular resolution (ICARUS, HELLAZ), the neutrino energy $E$ can be calculated from the measured energy $T_e$ of the recoil electron and the scattering angle $\theta$ between the neutrino and the recoil electron directions with the simple kinematical relation

$$E = \frac{m_e T_e}{p_e \cos\theta - T_e} \tag{A.1}$$

where $p_e$ is the momentum of the recoil electron. However, Cherenkov detectors (S-K and others) have a bad angular resolution due to random multiple Coulomb scattering of the recoil electron in water. Nevertheless, if the detector has a good energy resolution it is still possible to obtain the flux as a function of energy from the electron spectrum $n^{\rm ES}(T_e)$.

First of all, the ES event rate due to $\nu_e$ will be known from the SNO CC differential event rate:

$$n^{\rm ES}_{\nu_e}(T_e) = \int_{E_{\rm m}(T_e)} \frac{{\rm d}\sigma_{\nu_e e}}{{\rm d}T_e}(E,T_e)\,\phi_{\nu_e}(E)\,{\rm d}E \tag{A.2}$$

Hence, it will be possible to subtract $n^{\rm ES}_{\nu_e}(T_e)$ from the electron spectrum $n^{\rm ES}(T_e)$ in order to obtain the contribution $n^{\rm ES}_{\nu_\mu}(T_e)$ of $\nu_\mu$ and/or $\nu_\tau$ to the ES event rate: $n^{\rm ES}_{\nu_\mu}(T_e) = n^{\rm ES}(T_e) - n^{\rm ES}_{\nu_e}(T_e)$. Therefore,

$$n^{\rm ES}_{\nu_\mu}(T_e) = \int_{E_{\rm m}(T_e)} \frac{{\rm d}\sigma_{\nu_\mu e}}{{\rm d}T_e}(E,T_e) \sum_{\ell=\mu,\tau} \phi_{\nu_\ell}(E)\,{\rm d}E \tag{A.3}$$

is a measurable quantity. By differentiation we obtain

$$\begin{aligned} \frac{{\rm d}n^{\rm ES}_{\nu_\mu}}{{\rm d}T_e}(T_e) = & -\frac{{\rm d}E_{\rm m}}{{\rm d}T_e}(T_e)\,\frac{{\rm d}\sigma_{\nu_\mu e}}{{\rm d}T_e}(E_{\rm m}(T_e),T_e) \sum_{\ell=\mu,\tau} \phi_{\nu_\ell}(E_{\rm m}(T_e)) \\ & + \int_{E_{\rm m}(T_e)} \frac{{\rm d}^2\sigma_{\nu_\mu e}}{{\rm d}T_e^2}(E,T_e) \sum_{\ell=\mu,\tau} \phi_{\nu_\ell}(E)\,{\rm d}E \end{aligned} \tag{A.4}$$

with

$$\frac{{\rm d}E_{\rm m}}{{\rm d}T_e}(T_e^{\rm m}(E)) = 1 + \frac{m_e^2}{4E\,(E+m_e)} \tag{A.5}$$

$$\frac{{\rm d}^2\sigma_{\nu_\mu e}}{{\rm d}T_e^2}(E,T_e) = -\frac{4\,G_{\rm F}^2\,m_e}{\pi\,E}\left[\left(g_R^{\nu_\mu}\right)^2\left(1-\frac{T_e}{E}\right) + g_L^{\nu_\mu}\,g_R^{\nu_\mu}\,\frac{m_e}{2E}\right] \tag{A.6}$$

Notice that $\dfrac{{\rm d}^2\sigma_{\nu_\mu e}}{{\rm d}T_e^2}$ is suppressed at $E \simeq T_e$ and at large $E$. If we neglect the second term on the r.h.s. of Eq.(A.4), we obtain

$$\sum_{\ell=\mu,\tau} \phi_{\nu_\ell}(E) \simeq -\left[\frac{\dfrac{{\rm d}T_e^{\rm m}}{{\rm d}E}(E)}{\dfrac{{\rm d}\sigma_{\nu_\mu e}}{{\rm d}T_e}(E,T_e^{\rm m}(E))}\right] \frac{{\rm d}n^{\rm ES}_{\nu_\mu}}{{\rm d}T_e}(T_e^{\rm m}(E)) \tag{A.7}$$

with

$$T_e^{\rm m}(E) = \frac{2E^2}{m_e+2E} \tag{A.8}$$

$$\frac{{\rm d}T_e^{\rm m}}{{\rm d}E}(E) = \frac{4E\,(E+m_e)}{(2E+m_e)^2} \tag{A.9}$$

$$\frac{{\rm d}\sigma_{\nu_\mu e}}{{\rm d}T_e}(E,T_e^{\rm m}(E)) = \frac{2\,{G_{\rm F}}^2\,m_e}{\pi}\left(g_L^{\nu_\mu} - g_R^{\nu_\mu}\frac{m_e}{m_e+2E}\right)^2 \tag{A.10}$$

Once $\sum_{\ell=\mu,\tau} \phi_{\nu_\ell}(T_e)$ is derived with Eq.(A.7), it is necessary to estimate the second term in the r.h.s. of Eq.(A.4). If it is found to be much smaller than the first term, then the value of $\sum_{\ell=\mu,\tau} \phi_{\nu_\ell}(T_e)$ derived from Eq.(A.7) can be used as a starting point for a numerical solution of Eq.(A.4):

$$\begin{aligned} \sum_{\ell=\mu,\tau} \phi_{\nu_\ell}(E) = & - \left[ \frac{\frac{\mathrm{d}T_e^{\mathrm{m}}}{\mathrm{d}E}(E)}{\frac{\mathrm{d}\sigma_{\nu_\mu e}}{\mathrm{d}T_e}(E, T_e^{\mathrm{m}}(E))} \right] \left\{ \frac{\mathrm{d}n_{\nu_\mu}^{\mathrm{ES}}}{\mathrm{d}T_e}(T_e^{\mathrm{m}}(E)) + \right. \\ & + \sum_{k=1}^{\infty} \int_E \mathrm{d}E_1 \, \frac{\mathrm{d}^2\sigma_{\nu_\mu e}}{\mathrm{d}T_e^2}(E_1, T_e^{\mathrm{m}}(E)) \left[ \frac{\frac{\mathrm{d}T_e^{\mathrm{m}}}{\mathrm{d}E}(E_1)}{\frac{\mathrm{d}\sigma_{\nu_\mu e}}{\mathrm{d}T_e}(E_1, T_e^{\mathrm{m}}(E_1))} \right] \int_{E_1} \mathrm{d}E_2 \ldots \\ & \left. \ldots \int_{E_{k-1}} \mathrm{d}E_k \, \frac{\mathrm{d}^2\sigma_{\nu_\mu e}}{\mathrm{d}T_e^2}(E_k, T_e^{\mathrm{m}}(E_{k-1})) \left[ \frac{\frac{\mathrm{d}T_e^{\mathrm{m}}}{\mathrm{d}E}(E_k)}{\frac{\mathrm{d}\sigma_{\nu_\mu e}}{\mathrm{d}T_e}(E_k, T_e^{\mathrm{m}}(E_k))} \right] \frac{\mathrm{d}n_{\nu_\mu}^{\mathrm{ES}}}{\mathrm{d}T_e}(T_e^{\mathrm{m}}(E_k)) \right\} \end{aligned} \tag{A.11}$$

| $\nu_e \to \nu_S$ | | | | Lower Bound for | |
|---|---|---|---|---|---|
| | $\Delta m^2\,(\mathrm{eV}^2)$ | $\sin^2 2\vartheta$ | $\mathrm{P}^{\max}_{\nu_e\to\nu_e}$ | $\langle \mathrm{P}_{\nu_e\to\nu_S}\rangle_{\mathrm{NC}}$ | $\langle \mathrm{P}_{\nu_e\to\nu_S}\rangle_{\mathrm{ES}}$ |
| VACUUM OSC. | $6.3\times10^{-11}$ | 0.77 | 0.61 | 0.43 | 0.45 |
| SMALL MIX MSW | $4.5\times10^{-6}$ | $7.0\times10^{-3}$ | 0.58 | 0.30 | 0.27 |

Table 1. Results of the calculation of $\mathrm{P}^{\max}_{\nu_e\to\nu_e}$ (in the energy range that will be explored by SNO with $T_e^{\mathrm{th}} = 4.5\,\mathrm{MeV}$) and $\langle \mathrm{P}_{\nu_e\to\nu_S}\rangle_{\mathrm{NC;ES}}$ in a model with $\nu_e$–$\nu_S$ mixing. The values of $\Delta m^2$ and $\sin^2 2\vartheta$ used are also given. These values were obtained in refs.[19, 20] from the analysis of the existing experimental data.

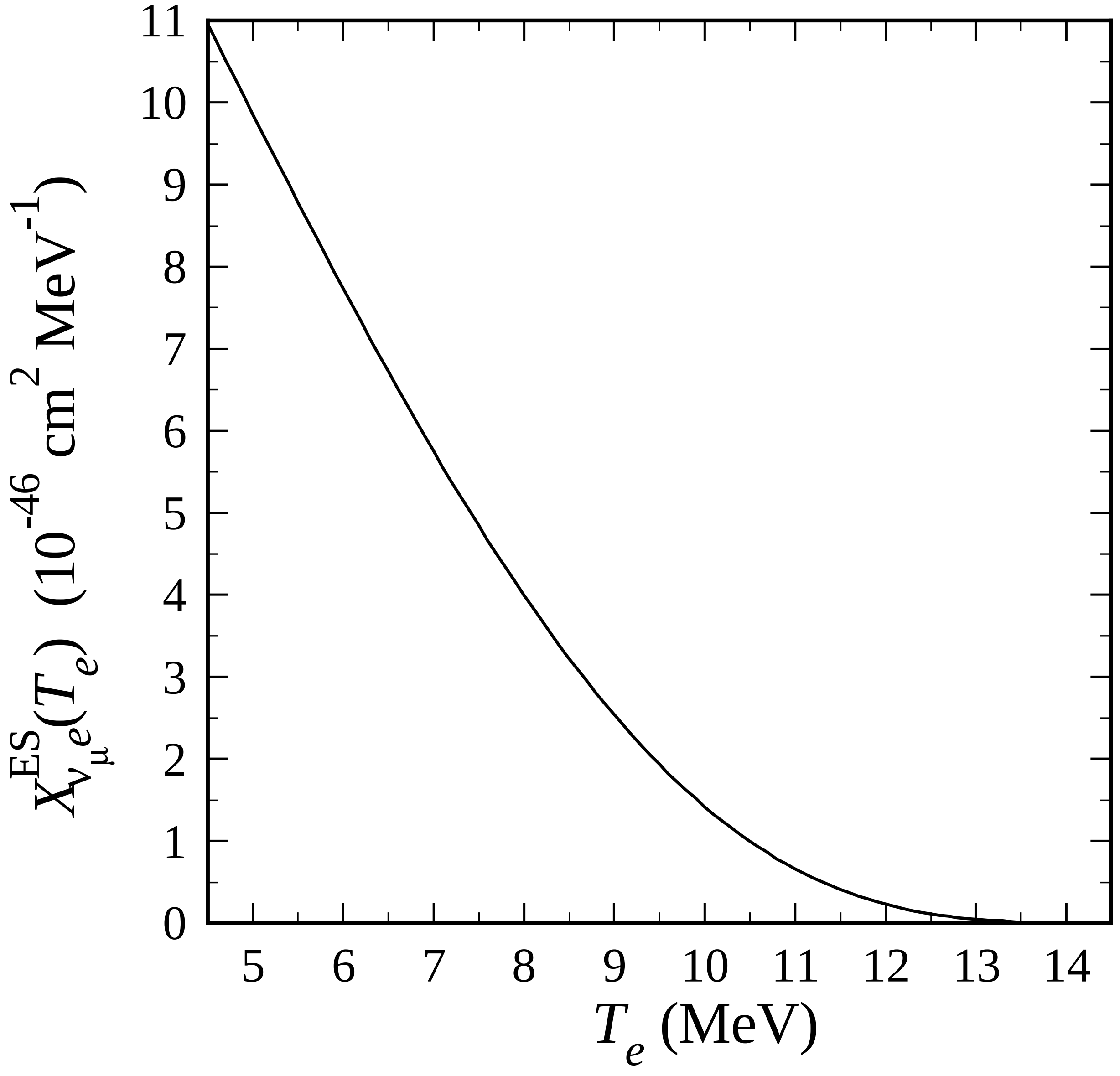

Figure 1. Plot of the function $X^{\rm ES}_{\nu_\mu e}(T_e)$ (see Eq.(2.11)). $T_e$ is the kinetic energy of the recoil electron in the ES process. The depicted energy range will be explored by SNO with $T_e^{\rm th} = 4.5\,\mathrm{MeV}$.

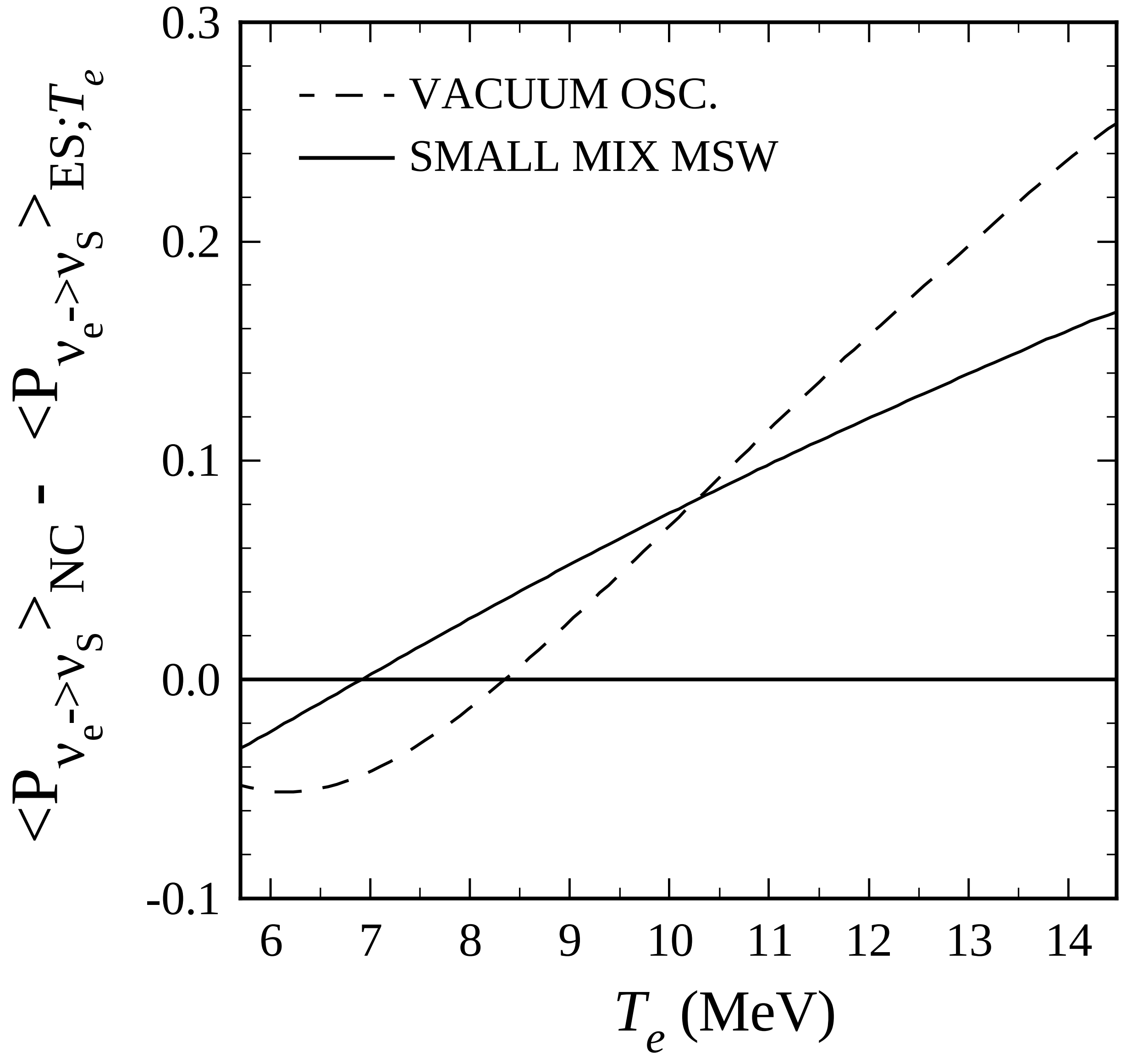

Figure 2. Results of the calculation of the right-hand side of Eq.(3.4) divided by the initial $^8$B $\nu_e$ flux $\Phi_{\nu_e}^{^8\mathrm{B}}$ ($T_e$ is the kinetic energy of the recoil electron in the ES process) in a model with mixing of $\nu_e$ and $\nu_{\mathrm{S}}$. The two curves correspond to vacuum oscillations and small mixing angle MSW transitions. The values of the parameters $\Delta m^2$ and $\sin^2 2\vartheta$ used in the calculation are given in Table 1.

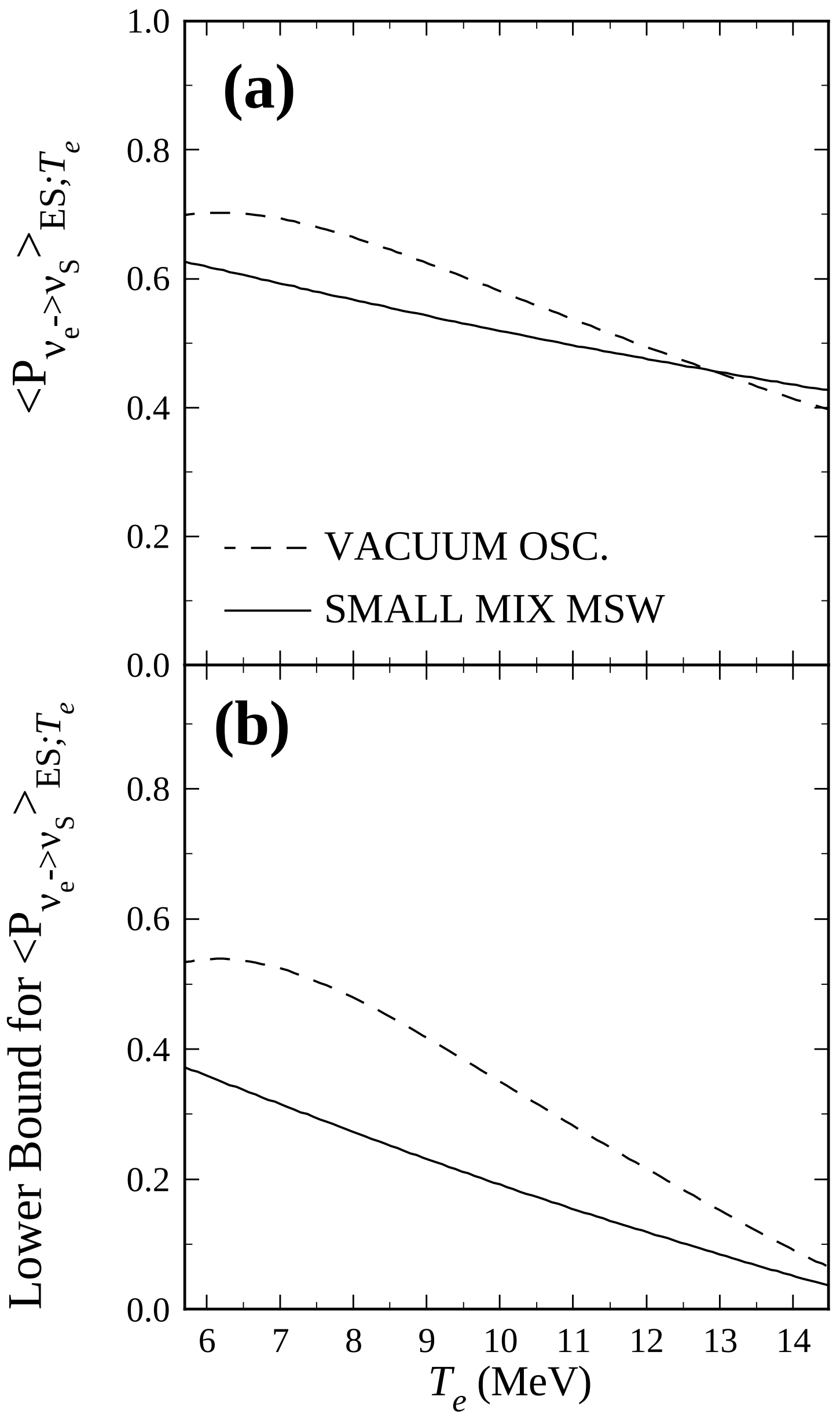

Figure 3. (a) Transition probability $\langle \mathrm{P}_{\nu_e \to \nu_S} \rangle_{\mathrm{ES};T_e}$ as a function of $T_e$ calculated in a model with mixing of $\nu_e$ and $\nu_S$. (b) Corresponding lower bound for $\langle \mathrm{P}_{\nu_e \to \nu_S} \rangle_{\mathrm{ES};T_e}$. The two curves correspond to the cases of vacuum oscillations and small mixing angle MSW transitions. The values of the parameters $\Delta m^2$ and $\sin^2 2\vartheta$ used in the calculation are given in Table 1.

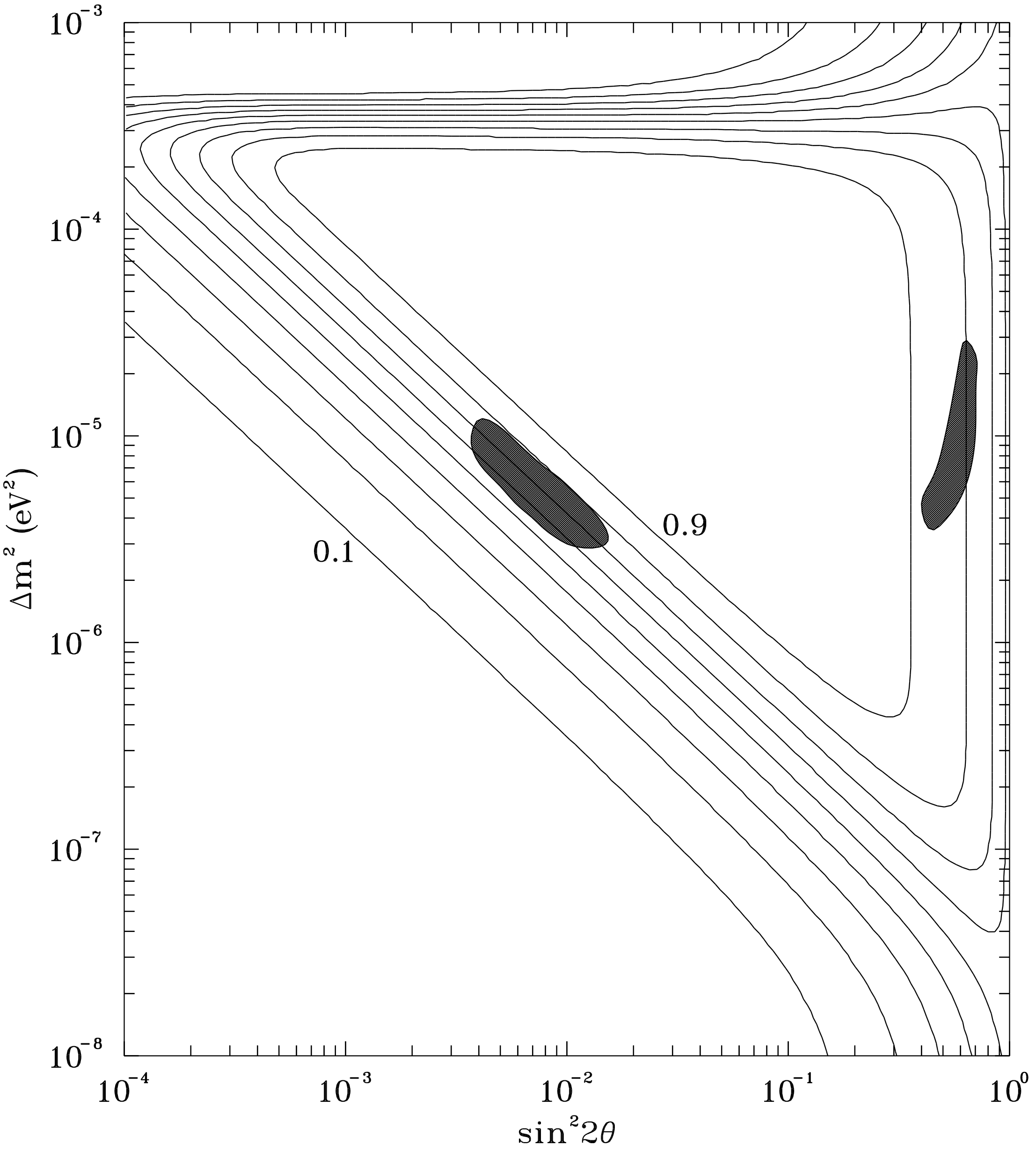

Figure 4. The ratio $R^{\rm NC}$, which characterizes the relative contribution of $\nu_\mu$ and/or $\nu_\tau$ to the rate of NC events. The curves correspond to the values of $R^{\rm NC}$ equal to 0.1, 0.2, ..., 0.9. The two regions of values of $\Delta m^2$ and $\sin^2 2\vartheta$ which were obtained from the analysis of the existing data [4] are also plotted.

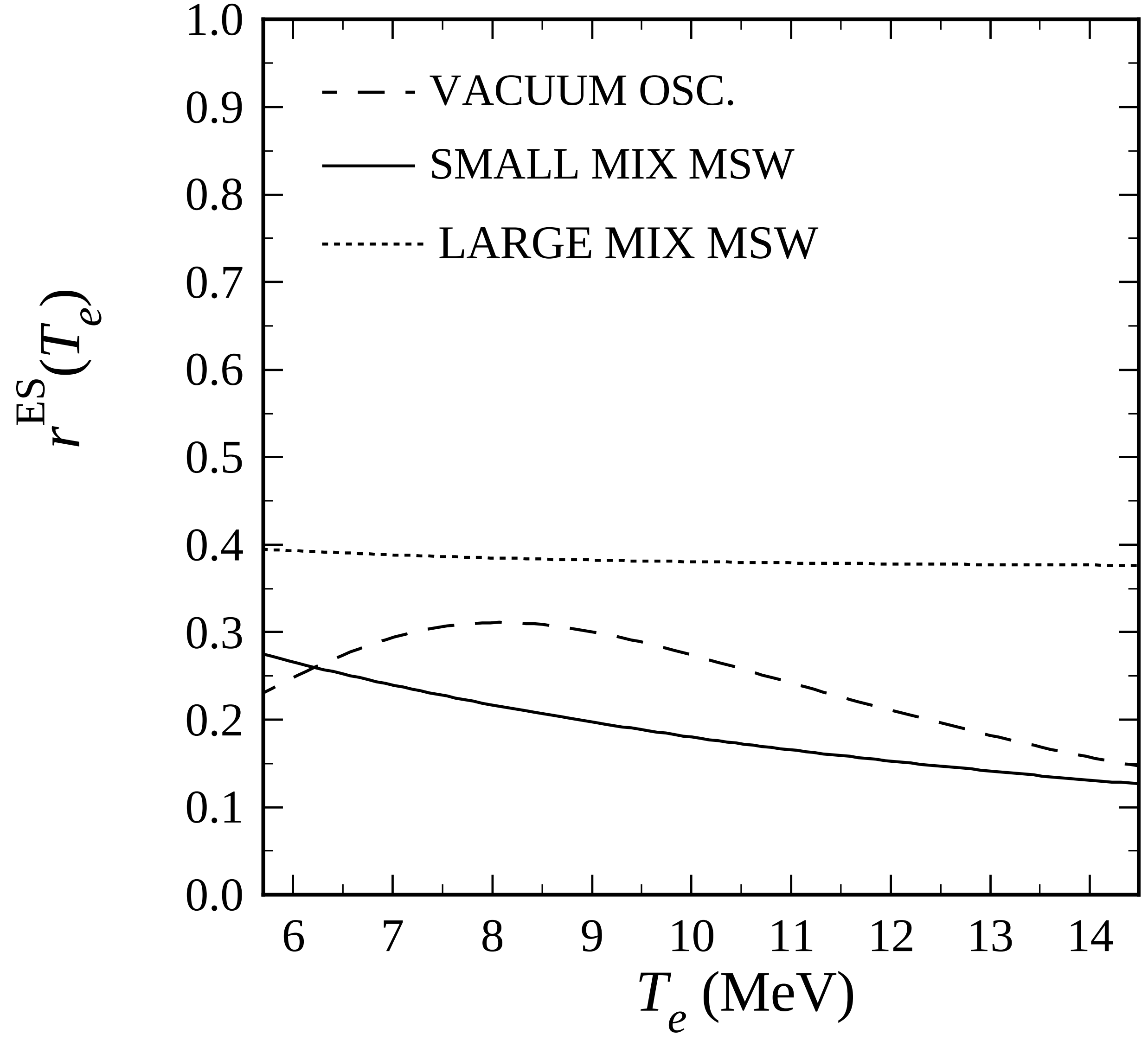

Figure 5. Relative contribution of $\nu_\mu$ (or $\nu_\tau$) to the ES event rate as a function of $T_e$ calculated in a model with $\nu_e$–$\nu_\mu$ (or $\nu_e$–$\nu_\tau$) mixing. The three curves correspond to the following transitions: vacuum oscillations with $\Delta m^2 = 7.8 \times 10^{-11}\,\mathrm{eV}^2$ and $\sin^2 2\vartheta = 0.78$; small mixing angle MSW transitions with $\Delta m^2 = 6.1 \times 10^{-6}\,\mathrm{eV}^2$ and $\sin^2 2\vartheta = 6.5 \times 10^{-3}$; large mixing angle MSW transitions with $\Delta m^2 = 9.4 \times 10^{-6}\,\mathrm{eV}^2$ and $\sin^2 2\vartheta = 0.62$. These values were obtained in refs.[19, 20] from the analysis of the existing experimental data.